\documentclass[sigconf]{acmart}

 \usepackage{xcolor}

\usepackage{tcolorbox}
\usepackage[subtle, tracking=normal]{savetrees}

\tcbuselibrary{skins, breakable}
\newtcolorbox{custombox}[1][]{%
    breakable,
    colback=orange!10,    
    colframe=orange!80!black, 
    coltitle=black,       
    fonttitle=\bfseries,  
    enhanced,
    rounded corners,      
    attach boxed title to top left={xshift=2mm,yshift=-2mm},
    boxed title style={colback=orange!30, colframe=orange!80!black},
    title=#1,             
}

\setcopyright{acmlicensed}
\copyrightyear{2025}
\acmYear{2025}
\acmDOI{XXXXXXX.XXXXXXX}
\acmConference[Conference acronym 'XX]{Make sure to enter the correct
  conference title from your rights confirmation emai}{June 03--05,
  2018}{Woodstock, NY}

\acmISBN{978-1-4503-XXXX-X/18/06}

\begin{document}


\title[HCI Challenges and Opportunities in Interactive Multi-Agentic Systems]{From Conversation to Orchestration: HCI Challenges \\ and Opportunities in Interactive Multi-Agentic Systems}



\author{Sarah Sch\"{o}mbs}
\affiliation{%
  \department{School of Computing and Information Systems}
  \institution{University of Melbourne}
  \city{Melbourne}
  \state{VIC}
  \country{Australia}}
\email{sschombs@student.unimelb.edu.au}
\orcid{0009-0001-3251-3199}

\author{Yan Zhang}
\affiliation{%
  \department{School of Computing and Information Systems}
  \institution{University of Melbourne}
  \city{Melbourne}
  \state{VIC}
  \country{Australia}}
\email{yan.zhang.1@unimelb.edu.au}
\orcid{0000-0003-2142-5094}

\author{Jorge Goncalves}
\affiliation{%
  \department{School of Computing and Information Systems}
  \institution{University of Melbourne}
  \city{Melbourne}
  \state{VIC}
  \country{Australia}}
\email{jorge.goncalves@unimelb.edu.au}
\orcid{0000-0002-0117-0322}

\author{Wafa Johal}
\affiliation{%
  \department{School of Computing and \\Information Systems}
  \institution{University of Melbourne}
  \city{Melbourne}
  \state{VIC}
  \country{Australia}}
\email{wafa.johal@unimelb.edu.au}
\orcid{0000-0001-9118-0454}

\renewcommand{\shortauthors}{Trovato et al.}

\begin{abstract}
Recent advances in multi-agentic systems (e.g. AutoGen, OpenAI Swarm) allow users to interact with a group of specialised AI agents rather than a single general-purpose agent. Despite the promise of this new paradigm, the HCI community has yet to fully examine the opportunities, risks, and user-centred challenges it introduces. We contribute to research on multi-agentic systems by exploring their architectures and key features through a human-centred lens. While literature and use cases remain limited, we build on existing tools and frameworks available to developers to identify a set of overarching challenges, e.g. orchestration and conflict resolution, that can guide future research in HCI. We illustrate these challenges through examples, offer potential design considerations, and provide research opportunities to spark interdisciplinary conversation. Our work lays the groundwork for future exploration and offers a research agenda focused on user-centred design in multi-agentic systems.
\end{abstract}


\keywords{agentic systems, multi-agent system, human-agentic interaction}



\maketitle

\section{Introduction}
Within just a few months of release, Large Language Models (LLMs) like ChatGPT have attracted millions of users \cite{milmo_chatgpt_2023}. Since then, foundation models have rapidly evolved and have shifted the focus from AI as decision-making tools to \textbf{agentic} systems capable of serving general-purpose roles in everyday life, reviving interest in Agent-Oriented Programming \cite{shoham1993agent} along the way. Recent advances, e.g. from Microsoft \cite{wu2023autogen} and OpenAI \cite{noauthor_openai-agent_2025}, now go even further and introduce frameworks for multi-agentic systems that introduce the idea of users not only interacting with one AI agent, but handing over tasks to multiple AI agents. This paradigm shift in human-AI interaction, driven by advances in LLMs, represents a significant departure from the `traditional' model of relying on a single, static AI agent. Instead, it re-imagines human-AI interaction as the gateway to multiple specialised AI agents, each designed to perform distinct roles within a collective system. In this model, individual agents are uniquely tailored and prompted to achieve specific goals, execute tasks and access tools (e.g. APIs). Importantly, these agents are not siloed; they can communicate, exchange information, and collaboratively solve problems, which contributes to emergent complexities end-users must navigate. The idea of multi-agent systems (MAS) is not new and has traditionally been considered in fields such as robotics, for example, with the introduction of swarms \cite{mataric_issues_1995,tessier_conflicting_2005}. However, multi-\textbf{agentic} systems, backed by foundation models, are expected to be accessible to the general public. This raises important questions regarding their perceptual and behavioural effects on end-users, as well as the design challenges associated with creating interfaces that support effective interaction with such systems. 



The implications of this paradigm extend far beyond efficiency and technical convenience. By structuring software around a network of interactive agents, multi-agentic systems open the door to greater personalisation and adaptability in user experiences. While current advances predominantly remain on the developer side, we are beginning to see early steps toward making these capabilities accessible to end-users. For example, recent work on Magentic UI \cite{fourney_magentic-one_2024} introduces a prototype chat user interface that allows end-users to delegate tasks to a multi-agentic team called Magentice-One \cite{fourney2024magentic}. Magentic-One \cite{fourney_magentic-one_2024} consists of a hierarchical ``generalist'' multi-agentic team with a top level supervisor agent, i.e. ``orchestrator agent'', and a set of specialised subagents (e.g. WebSurfer, Coder, FileSurfer) capable of completing open-ended general purpose tasks (e.g. browsing the web or executing relevant Python code) in real-time. In this system architecture, the supervisor agent interprets the user's prompt, plans the task, delegates to the specialised agents, and monitors the task progress. 

These developments hint at a near future in which multi-agentic systems are programmable and usable by a broader audience. While the opportunities are vast, this approach also introduces new challenges in terms of transparency and interpretability of multi-agentic systems, as non-technical end-users must navigate not only the outcomes but also the interactions between agents that produce them. Furthermore, ensuring human control over such systems, particularly when emergent behaviours lead to unexpected outcomes, raises critical questions about trust and risks. These dynamics highlight the need for an in-depth investigation into how multi-agentic systems reshape our relationship and interactions with AI from an end-user's perspective. Addressing these challenges is crucial to ensure a responsible design and deployment of multi-agentic systems for the general public. For instance, creating mechanisms that allow users to easily trace and understand inter-agent communications will be essential for maintaining transparency and mitigating risks. Similarly, developing safeguards to prevent cascading errors or rogue behaviours among agents will be crucial to ensure reliability and user safety \cite{chan_visibility_2024}. 

Importantly, multi-agentic systems can take on various architectures, ranging from a multi-body to a fully decentralised agent architecture. This paper focuses on the hierarchical architecture, a multi-agentic system design in which, in its simplest form, a supervisor agent acts as the central interface between the user and a set of specialised sub-agents, similar to Magentic-One \cite{fourney_magentic-one_2024}. Our rationale for focusing on this architecture is motivated by two key factors: First, it reduces the cognitive burden on the end-user by abstracting away the complexity of interacting with multiple autonomous agents. Instead of managing a distributed system directly, which can be cognitively demanding (see early works in swarm robotics e.g. \cite{lee_emerging_2001, wong_workload_2017}), a supervisor agent can plan, delegate, and coordinate tasks internally. Second, the hierarchical architecture is emerging as a popular architecture in current prototype implementations/applications \cite{fourney_magentic-one_2024} or is often proposed as the default structure in current frameworks \cite{wu_autogen_2023}. Given its growing adoption and practical promise, it is essential to investigate the design challenges that arise specifically for end-users when interacting with systems built on this architecture. The central research question in this work is therefore:

\begin{quote}
\textit{What are the design challenges and opportunities for end-user interaction introduced by a hierarchical architecture in multi-agentic systems?}
\end{quote}

To explore this research question, we examine alternative multi-agent architectures to contextualise and contrast the supervisor paradigm, as well as compare to current single-agentic systems. This comparative perspective allows us to more precisely articulate the design challenges (including trade-offs and constraints), and opportunities that define the hierarchical architecture.

We contribute to an important, timely and growing body of research on agentic systems. First, we explore agentic architectures and describe their principles and key features from a human-centred perspective. While literature and use cases are still sparse, we build on this knowledge and our expertise, and use the characteristics of the current tools already available for developers to identify a set of design challenges that can be used to drive future research in Human-Multi-Agent interaction. We connect these challenges to broader HCI challenges, illustrate them through examples and explore how they intersect with opportunities and risks. Recognising these risks, we advocate for proactively anticipating potential harms in the design of user interfaces for multi-agentic systems, following \citet{chan_harms_2023}. 
Our work addresses the pressing need for the HCI community to engage with the rapid evolution of single-LLMs to multi-agentic architectures through the end-user lens. 

\section{From Agent to Swarm}

\subsection{Agentic and System Architectures}
\label{sec:architecture}

Broadly defined, an \textbf{agent} can perceive its environment using sensors and act on it through effectors~\cite{russell1995modern}, and \textbf{agency} denotes ``the exercise or manifestation of this capacity''~\cite{zalta1995stanford}. With advancements in AI, agents now embody key properties that determine their interaction with humans, other agents, and the environment, such as autonomy~\cite{castelfranchi1994guarantees} and social ability~\cite{castelfranchi1994guarantees}, reactivity, proactiveness, as well as additional traits typically associated with humans, including mentalistic reasoning~\cite{shoham1993agent}, emotional capacity~\cite{bates1994role}, and anthropomorphic features~\cite{maes1995agents}. In line with \citet{chan_harms_2023}, we define \textbf{agentic systems} as systems that include AI agents with a high degree of agency, characterised by underspecification, directness of impact, goal-directness, and long-term planning, as seen in examples like ChemCrow~\cite{m2024augmenting} and ROSAnnotator~\cite{zhang2025rosannotator}. While agentic systems with a single agent have shown impressive capabilities, they face limitations as tasks grow more complex. They often struggle with hallucinations, tool management, and representing diverse perspectives~\cite{jiang2024multi}. As a result, both academia and industry are turning to systems with multiple agents.

Traditionally, multi-agent systems (MAS) have been investigated in distributed computing in the context of information agents~\cite{haverkamp1998intelligent} and in robotics in the context of swarms~\cite{ota2006multi}. Past work defined MAS as ``a collection of, possibly heterogeneous, computational entities, having their own problem-solving capabilities and which are able to interact in order to reach an overall goal''~\cite{oliveira1999multi} or as ``multiple agents collaborating to solve a complex task''~\cite{dorri2018multi}. In our work, we define \textbf{multi-agentic systems} as systems composed of multiple AI agents, each exhibiting a high degree of agency as defined above, exemplified by Metagpt~\cite{hong2023metagpt} and AutoGen~\cite{wu2023autogen}.

\begin{figure}
    \centering
    \includegraphics[width=1\linewidth]{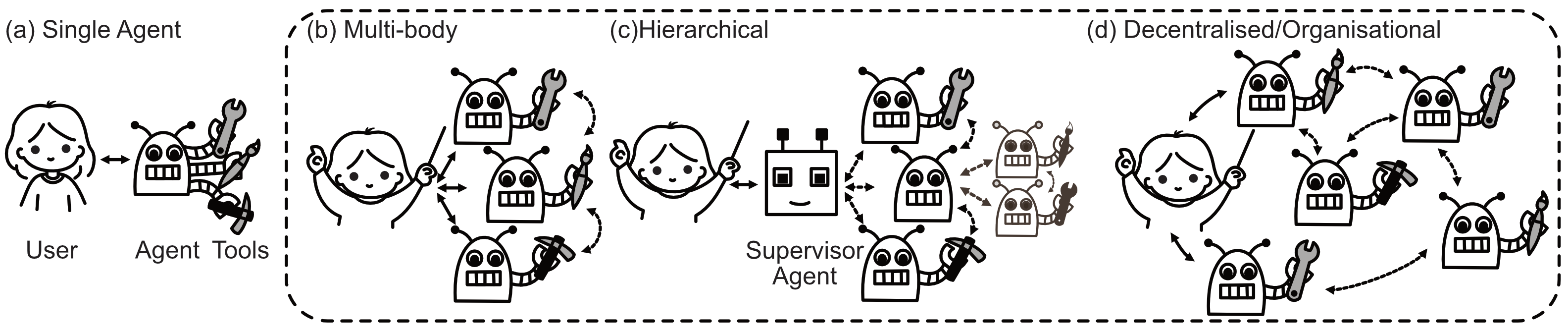}
    \caption{This figure illustrates four agentic system architectures. The first one is a single-agent system and the others are multi-agentic systems.}
    \vspace{-20pt}
    \label{fig:archi}
\end{figure}

In general, the architectures of agentic systems can be classified into four types: single agent, multi-body, hierarchical, and decentralised/organisational. The latter three fall under the umbrella of multi-agentic systems, which each offer distinct mechanisms for interaction, task allocation, and coordination, tailored to address varying complexities and operational demands. \autoref{fig:archi} (a) illustrates a typical single-agent system. In this architecture, the user directly interacts with a single agent that accesses and selects multiple tools (e.g. APIs) to fulfill requests, and generates responses. However, single-agent systems often lack efficiency and robustness for complex tasks~\cite{wu2023autogen}. The multi-body system (\autoref{fig:archi} (b)) enables users to assign roles and sub-tasks to multiple specialised agents, each focusing on domain-specific expertise. Agents communicate internally to share information, coordinate actions, and integrate outputs, leveraging parallel processing to speed up task completion. While effective for tasks requiring diverse perspectives, this architecture can potentially increase the user's cognitive burden \cite{zang_large_2024}. The hierarchical architecture (\autoref{fig:archi} (c)) features a primary agent that interacts with the user, decomposes tasks into sub-tasks, and delegates them to specialised subordinate agents. The primary agent `oversees' progress, coordinates efforts, and integrates outputs, which enhances robustness~\cite{gao2024agentscope} and may reduce users' cognitive load. This structure is implemented in OpenAI Agent~\cite{noauthor_openai-agent_2025}, where a triage agent serves as the primary agent, delegating tasks to other specialised agents. Another example is AutoGen~\cite{wu2023autogen}, which employs a conversable agent as the top-level agent, supported by subclasses agents such as the Assistant Agent, User Proxy Agent, and Group Chat Agent. In the decentralised/organisational system (\autoref{fig:archi} (d)), users and agents share task responsibilities in a fluid, collaborative environment with peer-to-peer communication for real-time coordination. This non-hierarchical structure enables a high degree of interconnectedness, which allows users to customise and group agents while facilitating direct, peer-to-peer interactions without a central supervising entity.

\subsection{Multi-Agentic Systems in Practice}



With the commercialisation of LLMs, researchers have investigated users' interaction modes with these systems. These interaction modes can be categorised into four distinct types~\cite{gao2024taxonomy}. Firstly, prompting represents the most commonly used interaction method, which includes text-based conversation~\cite{zhu2023leveraging} and prompting with reasoning~\cite{fu2022complexity}. An example of the latter is the Chain-of-Thought (COT) approach~\cite{wei2022chain}. Secondly, users interact with LLMs through thoughtfully designed user interfaces (UIs). Such interfaces assist users by facilitating structured prompt inputs~\cite{dang2023choice}, offering multiple output options~\cite{jiang2022discovering}, visualising iterative interactions~\cite{brade2023promptify}, enabling interaction testing~\cite{kim2023cells}, and providing transparent reasoning sequences~\cite{wu2022ai}. Thirdly, context-based interaction enables LLM systems to understand and incorporate contextual information. This includes explicit context, such as integrating context into prompts or providing pre-existing information~\cite{yuan2022wordcraft}, as well as implicit context, such as employing role play~\cite{kumar2023exploring} or example-based prompts~\cite{xiao2023supporting}. Finally, interaction with an agent facilitator enhances team performance. In this mode, a team engages with an LLM system, which facilitates workflows by aiding communication, supporting decision-making, and assigning tasks to appropriate team members~\cite{duan2021bridging, zheng2023competent, shi2023retrolens}.

Interactions with multi-agentic systems share similarities with the patterns mentioned above, but also exhibit critical differences. Table~\ref{tab:rw} summarises key features of six popular multi-agentic systems as examples. The architectures of these leading systems are generally flexible and can be built and customised by developer users to suit their needs. 
Users' control and interaction with multi-agentic systems vary across systems, apart from the common step of initial task specification. In some frameworks, humans can join the agent group chat at some points, reviewing and approving the next step or supplying extra information before the following agent executes. Most systems expose a log stream for observation and tracking, but the visualisation differs: some print plain text to the terminal, whereas others provide a graphical user interface (GUI) or integrate with third-party SDKs; AgentOps~\cite{noauthor_agentops_2025} is one example to monitor and visualise agent activity. We will expand the user-interface review in the remainder of this section.

Through programming, developer users can build and customise their multi-agentic systems, which is currently the most widely used interface. Some systems also provide GUIs that support two primary goals: enabling end-users to intuitively delegate tasks~\cite{potts_magentic-ui_2025}, and enabling users to build their own agentic team without coding \cite{dibia_autogen_2024} or IDEs \cite{noauthor_langchain-ailanggraph-studio_2025}. In addition, some GUIs offer node graphs to build teams and visualise architecture, such as LangGraph Studio~\cite{noauthor_langchain-ailanggraph-studio_2025}. LangGraph Studio \cite{noauthor_langchain-ailanggraph-studio_2025} functions as an IDE for agentic workflows based on the LangGraph framework \cite{noauthor_langgraph_nodate}. It provides a visual interface through which developers can design, debug, and modify agent graphs, inspect individual steps, and manage execution logic. It introduces `interrupts' through which developers can set human-in-the-loop interruptions, e.g. before its first use. 

\begin{table}[]
\caption{A summary of the key features of example multi-agentic systems.}
\label{tab:rw}
\resizebox{\columnwidth}{!}{
\begin{tabular}{lllll}
\hline
System & Target user & Interface & User control\\ 
\hline
LangGraph~\cite{noauthor_langgraph_nodate} &  Developer & Code, GUI & Observable log stream; set task; interrupt\\
CrewAI~\cite{noauthor_crewai_nodate} &  Developer & Code, GUI & Observable log stream; set task; CLI\\
AutoGen~\cite{wu2023autogen} & Developer & Code, GUI & Observable log stream; set task; interrupt\\
OpenAI Agents~\cite{noauthor_openai-agent_2025} &  Developer & Code & Set task; track stream\\
MetaGPT~\cite{hong2023metagpt} & Developer & Code & Set task; interrupt; CLI\\
Magentic-One~\cite{fourney2024magentic} & Prompter \& developer & Code, GUI & Set task; log events\\

\hline
\end{tabular}
}
\end{table}

CrewAI Studio \cite{noauthor_crewai-studio_2025} is a no-code, user-friendly interface for creating agents and building ``crews''. After an agent is configured with its tools, it can be assigned to multiple crews and tasks, each with a trackable status. Within a crew, users select agents from a drop-down list and assign tasks. At present, CrewAI Studio supports only sequential and hierarchical architectures; the latter requires a manager (orchestrator) agent. More customised or decentralised teams still require coding. The interface also allows connections to external knowledge sources. During execution, the studio prints a live log stream, visible as plain text in a terminal, and displays the final output in the GUI.

AutoGen Studio \cite{dibia_autogen_2024} is a low-code user interface that enables researchers and developers to design and test multi-agent workflows and agentic teams based on the AutoGen framework \cite{wu_autogen_2023}, using intuitive drag-and-drop representations and node graph visualisation on canva. The interface includes features to define agents, set prompts, assign models, and configure tool use. It introduces human oversight mechanisms: developers can set termination conditions, view the execution flow, and monitor the conversation flow between agents. As an example of the usage of the AutoGen framework, Magentic-One~\cite{fourney_magentic-one_2024} is a `generalist multi-agentic team for solving complex tasks', and it features Magentic-UI~\cite{potts_magentic-ui_2025} as its user interface. Unlike the abovementioned systems, which require users to build their own teams, Magentic-One provides a well-developed team that is ready to use. Magentic-UI shifts focus toward prompting end-users and provides a simplistic chat interface through which the end-user can instruct the multi-agentic team to accomplish a task. Based on the user prompt, the orchestrator agent provides a step-by-step plan for the task execution, which can be edited and which needs to be approved by the user. The orchestrator agent is the gateway between the user and the specialised subagents; the user can solely communicate with the orchestrator agent. The interface further offers a live view window that conveys e.g. the web search of the WebSurfer agent in real-time.

Whereas current efforts largely remain on the developers' side, we can already observe shifts towards interfaces that allow end-users to leverage multi-agentic teams as well as low-code interfaces for developers to design, build and customise default agentic systems. Thus we can easily imagine interfaces that combine the aforementoned, i.e. that will allow the general public to act on a spectrum of designer, developer and end-user of such multi-agentic systems, which introduces important design challenges we need to anticipate to safeguard appropriate use.  
\section{Design Challenges for Multi-Agentic Systems}

One of the key distinctions between single-agent systems like OpenAI ChatGPT\footnote{https://openai.com/chatgpt/overview/} and multi-agentic systems like AutoGen \cite{wu_autogen_2023} lies in the fundamentally different roles taken on by the user. In a single-agent system, the user typically interacts directly with one agent, often an LLM-based AI assistant, through a traditional chat interface, with a single agent responsible for handling all tasks, end-to-end. When a user interacts with a multi-agentic system, however, the user delegates tasks to several agents, with each potentially playing a specialised role, executing specific tasks, while accessing respective tools (e.g. one agent is specialised in data retrieval, another in analysis, another in scheduling). This fundamental difference shifts the user's role from the primary driver, issuing commands directly, iteratively and receiving feedback from one agentic source, to the user as `the composer', delegating and handing over task responsibilities and overseeing multiple agents at a high-level. In the hierarchical architecture, the \emph{orchestrator agent} often then translates the high-level objectives to the agentic-team and coordinates subtasks, similar to arranging sections, musicians and instruments in a musical piece to bring the composer's vision to life. In line with the user's new role as a composer of an agentic team, the end-user is suddenly less concerned with micro-managing, delegating and overseeing individual tasks accomplished by a single agent, but with high-level goals and strategic direction.

Next, we describe our identified six design challenges, which arise from the foundational shift in the user’s role when interacting with multi-agentic systems compared to single-agent systems. We offer design considerations, pose open questions, and provide directions for future research. We illustrate design challenges through an example scenario, adapted to highlight key aspects of each challenge.

\vspace{-5pt}
\begin{custombox}[Example Scenario]
Consider a user selling items on a popular online marketplace\footnote{This example is extrapolated from \citet{goyal_designing_2024}, which examines alignment dimensions of AI agents acting as surrogates for users.}, an ideal scenario for a personal multi-agentic system. Specialised agents could include a Sales Agent to manage listings and inquiries, a Negotiation Agent to take over for pricing discussions, a Refunds Agent to handle complaints and returns, and a Review Agent to manage feedback and address negative reviews in coordination with the Refunds Agent.
\end{custombox}


\subsection{DC1: Reducing Opaqueness in Agentic \\ Teamwork} 
\label{sec_dc1}
To simplify agent and task orchestration, the user offloads their orchestration responsibilities to a dedicated top-level supervisor agent, i.e. orchestrator agent, which we will be referring to as a Manager Agent in the following examples. This supervisor agent then acts as a gateway, managing interactions between the user and specialised agents. While this approach may simplify coordination and reduce cognitive load, it may also make the system opaque and introduce challenges related to \emph{transparency}, as the user may struggle to understand or influence the underlying agentic system. 

To account for the user's new role as the composer of an agentic team, we identified the challenge of designing high-level interfaces that enable users to transparently orchestrate multi-agentic systems, similar to a `control panel' or an `orchestration interface'. A high-level interface should allow users (a) to delegate high-level tasks to the orchestrator agent similar to Magentic-UI \cite{potts_magentic-ui_2025} which provides a simple chat interface to communicate with the orchestrator agent, but also (b) access lower-level scaffolding when needed, e.g. to assign roles, specify constraints or to monitor how agents are exchanging information among themselves. However, we acknowledge that providing an orchestration interface should \textit{support} human-multi-agent interaction, not overburden or hinder the user due to complexity. This balance between transparency, control and an appropriate level of information has been addressed as an overarching challenge in human-agentic conversation by \citet{bansal_challenges_2024}; highlighting that ``it can become counterproductive when instructing the agent becomes burdensome or reviewing agent outputs becomes overly cognitively taxing'', which only increases in complexity when multiple agents are involved. But \textbf{how can we design orchestration interfaces without a high degree of complexity, given the various agents involved and multitude of scaffolding needs?} 

Much like organisational systems, we could envision assigning agents into functional groups and designing an interface that supports user access and control at the group level. In the context of agent organisations, \citet{goos_role_2003} defines groups as ``a set of agents that are related via their roles, where these relationships must form a connected graph within the group.'' Designing interfaces to include agent grouping could reduce complexity and support users monitoring and safeguarding multi-agentic systems. Grouping is particularly intriguing when designing for hierarchical architectures with multiple supervisor agents each managing a distinct subset of agents; similar to teams, departments and divisions in an organisation. Groupings could visually distinguish the supervisor agent’s role from subordinate agents and agent groups, which could enable users to understand workflows, track delegation paths, and monitor tasks more effectively. 
One way in which an orchestration interface could visually convey the underlying hierarchical architecture of multi-agentic systems is through an organigram, which typically represents the structure of an organisation by illustrating relationships and hierarchies between different groups and roles. Other potential design approaches to support user oversight through an orchestration interface for hierarchical architectures could draw inspiration from organisational practices, such as implementing stand-up style check-ins with supervisor agents. For instance, the top-level orchestrator agent could act as the default primary point of contact for the user, similar to a General Manager Agent, while the interface still allows the user to initiate direct interactions with division-level supervisor agents when needed. 


\vspace{-5pt}
\begin{custombox}
To reduce their workload, a user designs a General Manager Agent to act as their `right hand', tasked with managing the team of specialised agents, such as the Sales Agent, Negotiation Agent, and Refunds Agent. 
\end{custombox}

\subsection{DC2: Interacting in Parallel}
Based on the sequential nature in which single-agent systems work, the user typically asks, the agent responds, and the user follows up. The sequential back and forth allows the user to iteratively fine-tune both their user input and the agent output through `traditional' prompting, and personalise LLMs~\cite{kirk_benefits_2024}. When interacting with multi-agentic systems, however, agents act and interact in parallel, as specialised sub-agents may not be `waiting' for user input requested by a supervisor agent or its handoff. The parallel nature of multi-agentic systems introduces challenges related to dependencies, synchronicity, but also user-agent conversation. It introduces the question: \textbf{How do we design interfaces that support the parallel nature of multi-agent systems?}. In current frameworks \cite{noauthor_openai-agent_2025,wu_autogen_2023,fourney_magentic-one_2024}, a top-level supervisor agent is often proposed as the primary communication channel between the user and the multi-agentic system. While the design simplifies interaction by abstracting internal complexity, it may obscure the parallel nature of the system as well as introduce practical challenges for the user. As systems may get more complex or decentralised, involving multiple supervisor agents, the user may need to communicate directly to specific supervisor agents, or may wish to update instructions or constraints through the orchestrator agent after delegation, while subagents execute tasks in real-time. In such cases, real-time user intervention may fail to take effect immediately, as the system continues executing tasks in parallel, which may introduce latency and delayed propagation of updates.

Interfaces like Magentic-UI and LangGraph Studio introduce mechanisms to support user control during parallel execution. In Magentic-UI, the user can pause execution by clicking into the live view window (e.g. while the WebSurfer Agent browses the web) to regain control before allowing the system to proceed. LangGraph Studio enables developers to insert explicit `interrupts' before or after node execution, such as before an agent calls a tool, to keep the human in the loop. While these mechanisms mark important first steps, they may not scale to more complex scenarios where multiple agents execute tasks concurrently, e.g. several WebSurfer agents running in parallel while others generate code, access different APIs in the background. Exploring communication and oversight challenges related to parallel processing involves asking in depth-questions such as \textbf{How to enable users to interact with multiple agents simultaneously without disrupting ongoing parallel processes?} and \textbf{How to allow users to prioritise or interrupt parallel tasks across agents when necessary, without losing control over the overall system?} User interfaces may need to display multiple `threads' of supervisor agent-user conversations or tasks concurrently to allow the user to interact with the respective supervisor agent(s) and to better represent the parallel nature of the system, a challenge we discuss under the broader goal of supporting appropriate mental models of multi-agent systems (see Section \ref{sec:mental_modal}).

\vspace{-5pt}
\begin{custombox}
A buyer requests a discount offer through the Supervisor Agent, which passes the request to the Negotiation Agent, while a Sentiment Agent simultaneously runs a sentiment analysis in the background to monitor and report conversations. Despite this concurrent activity, the user might only experience a single, cohesive interface, such as a chat window, through which interactions are managed. In this setup, the system obscures parallel operations.
\end{custombox}

\subsection{DC3: Designing for Emergent Complexity}
Another key aspect of multi-agentic systems is their emergent complexity, which we categorise into (1) structural and (2) operational emergent complexity. 
Structural emergent complexity relates to an evolving architecture and dynamic organisation of agents. First, with more agents come emergent behaviours that are harder to anticipate, since agents may dynamically influence each other, have different tool access, and could potentially create sub-agents the user is unaware of \cite{chan_visibility_2024}. Second, different system architectures involve different levels of interconnectedness and thus complexities, see \autoref{fig:archi}. Third, a system may evolve its architecture as tasks grow more complex, with emergent supervisor agents or sub-agents, workflow changes, or agents adapting roles and gaining new tool access. In the case of a hierarchical architecture, emergent complexity may manifest in two dimensions: growing width, where an orchestrator agent must coordinate an increasing number of sub-agents; and growing depth, where nested layers of supervisor and agent teams extend the hierarchy and challenge user oversight.

\vspace{-5pt}
\begin{custombox}
Over time, the Sales Agent directly requests the Review Agent to demand feedback from buyers after a sale, bypassing the Manager Agent and user. The Sales Agent also creates an Analysis Agent, a sub-agent, used by the Negotiation Agent for real-time market analysis to better negotiate ad-hoc requests and push sales.
\end{custombox}

Operational emergent complexity concerns complexity related to failure, misalignment, and cascading effects arising from multiple agents and their interconnectedness. \citet{chan_visibility_2024} states that ``sub-agent could be problematic because they introduce additional points of failure; each sub-agent may itself malfunction, be vulnerable to attack, or otherwise operate in a way contrary to the user’s intentions''. As discussed in \citet{fourney_magentic-one_2024}, failures and misalignment can happen on a small and large scale. An agent may accept website cookies without user oversight or attempts to log into a user account repeatedly, which results in a temporarily suspended user account. In addition, the (in)action or failure of a single agent can propagate throughout the system due to the interconnected nature of multi-agentic systems. This interdependence can lead to cascading failures, where one failure triggers a chain reaction, like a domino effect \cite{chan_visibility_2024}. The emergent complexity of multi-agentic systems may therefore amplify risks already existent in single-agent systems.

\vspace{-5pt}
\begin{custombox}
The Pricing Agent misinterprets a competitor's discount and erroneously adjusts prices, which triggers the Negotiation Agent to offer steeper discounts to current buyers, unaware of the error. Meanwhile, the Sales Agent initiates ads to aggressively promote the incorrectly discounted items. 
\end{custombox}

We present a set of challenges related to the emergent complexities of multi-agentic systems, not as a comprehensive list, but to initiate conversation within HCI and to guide future work in this space. To navigate structural emergent complexities, we re-emphasise the need to visualise agent hierarchies and to track and convey architecture changes (e.g. new emergent sub agents, new tool access). To navigate operational complexity, we propose to consider design solutions for interactions that allow users the following mechanisms: 1) debug, to track how decisions evolve(d) over time, 2) intervene, to step in when emergent behaviours lead to undesirable outcomes, 3) guide, to provide higher-level objectives and constraints that steer agent interactions or activates human control protocols. This is in line with \citet{terry_interactive_2024} who discuss the need to design interfaces for \textit{interactive} AI alignment that allows users to specify, verify and align on how the agent will produce the outcome while interacting with AI agents. More specifically, design efforts should explore \textbf{how to create user interfaces that prevent cascading failures}. \citet{chan_visibility_2024} discuss activity logs and real-time monitoring as key measures to improve visibility in agentic systems. While useful for debugging and post-accident tracking, we re-emphasise the need to design simple, user-friendly solutions that align with the emergent complexity of hierarchical multi-agentic systems. Design challenges related to emergent operational complexity further raise critical questions about how to effectively manage agent autonomy and support user situation awareness. As articulated by \citet{wong_workload_2017} in the context of multiple-robot supervision, interface and interaction design must account for decisions around autonomy models and the trade-offs between fixed and sliding autonomy, as well as the conditions under which control should transition between system and human. These ``automation triggers'' may be based on user input, system state, environment, or task-specific thresholds, and their definition and implementation in multi-agentic systems remain open design questions. 

\subsection{DC4: Resolving Conflicts}
In single-agent systems, conflicts usually occur between the user and the agent. There is no system internal negotiation, nor potential for internal conflicts. In a multi-agent system, however, conflicts too can arise between agent(s) and user, but also between agents themselves. Each agent can have partially overlapping or competing goals, which may lead to inter-agent negotiation or cooperation.


\vspace{-5pt}
\begin{custombox}
The Sales Agent, tasked with selling as much as possible as quickly as possible, initiates a significant price drop to secure a deal. Meanwhile, the Pricing Agent, focused on maintaining the price structure, rejects the discount. This creates a conflict: the Sales Agent prioritises speed and volume, while the Pricing Agent upholds financial stability. 
\end{custombox}


In hierarchical systems, conflict resolution between competing agent goals or diverging outcomes may fall to the top-level orchestrator agent. While this delegation may simplify coordination, it may undermine human oversight and reduce user control. This raises several design challenges and research directions: (1) how can we design interfaces to support user oversight in situations where agent-agent conflicts occur, (2) how to enable the user to take over control in agent-agent conflicts that are critical for the user and to resolve such conflicts, and (3) how to design mechanisms for the user to define clear boundaries within which an orchestrator is allowed to resolve conflicts to reduce the cognitive burden.

As illustrated, users may require transparency into `behind-the-scenes discussions'. One approach to this design challenge could be to design interfaces that enable users to observe agents negotiate the best solution or weigh alternative suggestions from multiple agents. New UI metaphors like `roundtable views' or `agent council dashboards' could reveal how agents discuss and settle on a final recommendation. This raises another crucial question, i.e. \textbf{who has the final say in multi-agent conflicts?}; a question closely tied to responsibility attribution. One might argue that, for meaningful human control \cite{cavalcante_siebert_meaningful_2023}, the user should always have the final say. However, this might result in continuous human involvement, diminishing the purpose of an orchestrator agent and multiple agents `working for you'. The other extreme would be to design a Mediator Agent responsible for resolving conflicts, while safeguarding the user's preferences and decisions involving ethical concerns, high stakes, or actions that significantly impact the user's goals or values. The latter raises the question of \textbf{whether different tools are needed for resolving conflicts with varying levels of harm} (e.g. reputational harm, financial loss, ethical harm). It also raises the question of \textbf{how much human oversight is required for successful conflict resolution in multi-agentic systems}, which highlights the delicate interplay between engagement alignment and agent autonomy, two important alignment dimensions when using AI agents as surrogates working on the user's behalf \cite{goyal_designing_2024}.

\subsection{DC5: Understanding Multi-Agentic Systems}
\label{sec:mental_modal}
An appropriate mental model is crucial for users to understand the capabilities, limitations, and to infer appropriate use (e.g. trusting information) when interacting with any type of algorithmic system. This is especially relevant given the probabilistic nature of these systems, which can produce unpredictable outputs \cite{cavalcante_siebert_meaningful_2023,prabhudesai_understanding_2023}, raising concerns in high-stakes applications (e.g. financial management), and contexts where the user does not have knowledge superiority \cite{pareek_trust_2024,wang_are_2021}. While these challenges are significant in single-agent systems, they become even more pronounced for multi-agentic systems: users may be required to (1) form a mental model of the agents within the system (roles, tasks, tools), (2) form multiple overlapping mental models corresponding to several agents within groups, and (3) form a holistic mental model of the system, and how agents interact collectively. \textbf{How can we design interfaces to support people's understanding of a multi-agent system and its agents involved?} \citet{chan_visibility_2024} discuss agent identifiers to increase the visibility of AI agents, i.e. ``agent cards''. Contrary to model cards for ML models \cite{mitchell_model_2019}, \citet{bansal_challenges_2024} encourage more simplified solutions to make agent information accessible and understandable. Notably, in multi-agentic systems, the involvement of multiple agents or specialised agent groups may make forming mental models cognitively demanding. \textbf{How can we design solutions that reduce cognitive load while still supporting an appropriate mental model of multiple agents simultaneously?} A more practical design solution could be the use of group cards, instead of individual agent cards. Such group cards could organise agents into functional categories and offer overarching tasks, goals, and access permissions. In some cases, even simple group labels might provide `enough' information for users to infer key details through shared attributes, similar to organisational settings, in which `the finance team' naturally suggest responsibilities related to financial operations, and permissions to handle sensitive data. 


Interestingly, Salesforce advertises their agents in Agentforce 2.0 \cite{agentforce_url)} as cartoonish, futuristic anthropomorphic or zoomorphic characters ready to work on your behalf. Similar conscious or unconscious design choices open up new research avenues around people's tendency to anthropomorphise \cite{waytz_mind_2014} multi-agentic systems, and anthropomorphism by design (e.g. framing, such as naming agents or giving them backstories \cite{darling_whos_2015,schombs_feeling_2023,roesler_anthropomorphic_2023}) and the effects on perceived competence, trustworthiness and reliance, but also risks related to biases (e.g. gender stereotypes) introduced by anthropomorphism. \citet{abercrombie_mirages_2023} discuss linguistic cues that induce anthropomorphism in dialogue systems and their associated harms. Related work shows that users are more likely to disclose information to servant agents than mentor agents in highly anthropomorphic contexts \cite{zhang_tools_2023}, which introduces considerations related to agent identity, and roles. \citet{kirk_benefits_2024} discuss that anthropomorphising personalised LLMs may lead to privacy risks by increasing users’ willingness to disclose sensitive information. Recent work by \citet{cohn_believing_2024} shows that users rate information provided by an LLM as less risky when communicated through first-person pronouns. However, while anthropomorphism is well-studied in HCI fields like Human-Robot Interaction (e.g. \cite{roesler_meta-analysis_2021,blut_understanding_2021}) and is emerging in the context of LLMs \cite{deshpande_anthropomorphization_2023}, controlled user studies are yet scarce for multi-agentic systems. We encourage future research to explore \textbf{how design factors in multi-agentic systems (e.g. framing, appearance) affect user interaction and perception, particularly regarding agent identity.}

\vspace{-5pt}
\begin{custombox}
The Sales Agent is named `Piggy' and portrayed as a cheerful expert in buyer interactions, while the Pricing Agent, `Paul,' is depicted as a strategist focused on pricing. Because of Piggy’s design, the user perceives Piggy as less capable for critical tasks, such as managing sensitive buyer data, and instead relies heavily on Paul, even for tasks outside the agent's role, like accessing account details, information Piggy was designed to handle securely.
\end{custombox}

\subsection{DC6: Navigating Trust and Explainability\\ in Multi-Agentic Systems}
Explainability and trust are key challenges in HCI that have gained much attention in the context of machine learning (ML) and AI-assisted decision-making systems. A growing body of literature on AI-assisted decision-making explores various types of explanations (e.g. uncertainty communication \cite{schombs_robot-assisted_2024,prabhudesai_understanding_2023}, cognitive forcing functions \cite{bucinca_trust_2021}, confidence scores \cite{zhang_effect_2020}, feature contribution \cite{wang_effects_2022}) to address over-reliance, support appropriate levels of trust, and recover trust post failure (e.g. \cite{pareek_trust_2024,vasconcelos_explanations_2023,wang_are_2021,zhang_effect_2020}). 
\citet{lee_trust_2004} described trust as ``an attitude that an agent will achieve an individual’s goal in a situation characterised by uncertainty and vulnerability.'' Unlike single-agent systems, where explainability focuses on a singular agent’s decision-making and corresponding trust calibration, multi-agentic systems require users to navigate varying levels of trust in different agents. This variability may challenge the design of explanations, as some agents may require greater transparency than others, depending on their role, task, tool access, or risks involved related to the agent or agent group; which may also inspire novel approaches to risk communication \cite{schombs_robot-assisted_2024,schombs_facevis_2024}. 


If the interface design allows the users to only interact with specific agents, e.g. the orchestrator agent, the perceived trustworthiness of the entire system may depend solely on that agent’s behaviour, communication style, and ability to surface relevant information. This raises open questions: \textbf{What happens when a failure occurs at the sub-agent level, but the supervisor agent abstracts, hides or ignores it?} \textbf{How does system trust evolve when errors arise at different layers of the hierarchy?} These questions point to design challenges around how and when to expose system internal workings or hidden processes, how to support trust calibration over time, and how to build recovery mechanisms. We encourage future research \textbf{to investigate the interplay between holistic system trust and trust in individual agents, the design of explanations that support appropriate levels of system reliance, and strategies for failure recovery and trust repair}. Beyond, multi-agentic systems may introduce new challenges to these concepts due to the high degree of agency involved \cite{chan_visibility_2024}.

\vspace{-5pt}
\begin{custombox}
The Pricing Agent sets prices too low, leading to financial loss, which causes the user to lose trust in the system even though the Sales Agent continues closing deals and the Review Agent keeps customers satisfied.
\end{custombox}

While this paper largely focuses on design challenges through the end-user's lens, we must address the inherent assumption that end-users bear the burden of operating across the developer-user spectrum. This raises considerable ethical questions and challenges. A user who both configures and deploys their own agentic team, which acts in open-ended environments, takes on responsibilities traditionally reserved for system designers. This shift raises ethical considerations around human control, responsibility, and accountability. In the next section, we discuss how HCI researchers can meaningfully contribute to this important research agenda.

\section{Opportunities for HCI researchers}
As users increasingly engage with systems composed of multiple autonomous agents, HCI researchers are uniquely positioned to shape how these interactions are designed, understood, and governed. Next, we outline key opportunity areas where HCI can contribute to building effective, transparent, and ethical multi-agentic systems.

\subsection{Understanding, Monitoring, and \\ Interacting with Multi-Agentic Systems}

An opportunity for HCI researchers lies in creating new ways to visualise agent-to-agent (A–A) communication. Multi-agentic systems operate through ongoing internal dialogues that often remain opaque to the user. By making these interactions visible, through techniques such as conversation graphs, timelines, and activity maps, interfaces can expose the `invisible work' occurring beneath the surface. Such transparency supports interpretability, alignment checking, and debugging, particularly as systems become more autonomous.

These shifts also call for a rethinking of interaction paradigms. Multi-agentic systems introduce users to many-to-one or even many-to-many communication dynamics, which necessitate new metaphors and models for collaboration. Interfaces inspired by team communication tools, such as channel-based coordination or direct agent messaging, may offer more scalable and intuitive patterns. However, these approaches also require careful design to manage agents' presence, to avoid users feeling outnumbered or overwhelmed by a crowd of semi-autonomous agents.

As the boundaries between users and systems become more fluid, designers must develop explainable, interruptible, and attention-aware interfaces. These interfaces should support selective transparency, notifying users of critical decisions or conflicts without requiring constant supervision. Visual and interaction mechanisms that enable layered understanding are essential to sustaining human agency within complex systems.

\subsection{Control, Delegation, and Oversight}
With increasingly autonomous and modular multi-agentic systems, the challenge of maintaining meaningful user control becomes more complex. Traditional paradigms of command-and-response interaction are no longer sufficient; instead, users must navigate a landscape in which agents initiate actions, coordinate with one another, and sometimes operate beyond the user’s immediate awareness. This raises critical questions about how to structure delegation, monitor autonomy, and intervene effectively.

One key opportunity for HCI research is to design new models of oversight that recognise and embrace the distributed nature of multi-agentic systems. Users must be able to delegate goals or responsibilities to the system while retaining an understanding of how decisions are made and when intervention is needed. Mechanisms such as delegation protocols, approval gates, and review checkpoints can help formalise the boundaries of agent initiative. 

Control in such systems is further complicated by asymmetries of knowledge and authority, both between users and agents, and among agents. Some agents may have privileged access to information or play orchestration roles, while others execute subtasks or serve as monitors. Interfaces must therefore not only make user–agent relationships visible, but also map intra-agent hierarchies and negotiation processes. This opens up possibilities for tools that allow users to visualise and adjust internal governance models, ranging from fully autonomous teams to tightly supervised collectives.

\subsection{User Experience and Ethical Implications}
As multi-agentic systems introduce new paradigms of distributed autonomy, users are increasingly placed in unfamiliar roles: not as direct operators of tools, but as participants in loosely coupled teams of agents. This shift has significant implications for user experience design, as well as for the ethical and psychological dimensions of interacting with AI teams.

One core challenge is the ``one-among-many'' effect: users may experience disorientation, alienation, or reduced agency when engaging with a group of seemingly coordinated agents \cite{podevijn2016investigating}. Traditional interfaces that emphasise direct manipulation or dialogue with a single AI entity may not scale well in these contexts. HCI must therefore explore how to foster a sense of control, clarity, and belonging when users are embedded in digital collectives \cite{shiomi2019number}.

A related concern is how users interpret intent, role, and responsibility in a system where decisions emerge from the coordination of many agents \cite{bejarano2023no}. Attribution becomes diffuse: was an outcome the result of a single agent’s error, a flaw in coordination, or a broader systemic issue? Understanding how users assign credit, blame, or trust within multi-agentic systems is essential to shaping transparent and resilient experiences.

As agency is distributed across autonomous systems, so too is accountability. HCI researchers are uniquely positioned to design interfaces that surface the ethical boundaries of system behaviour, indicating when agents are acting on user instructions, internal objectives, or emergent coordination. Additionally, mechanisms for signalling disagreement or uncertainty within the agent collective could play a critical role in maintaining user trust and moral clarity.

There is also a need to examine the social and psychological effects of prolonged exposure to agent teams. What are the emotional consequences of interacting with systems that exhibit internal relationships, disagreements, or alignment? How do users anthropomorphise, align with, or resist agent collectives? \cite{shiomi2019number} These questions call for empirical research that bridges interaction design with theories from social psychology, organisational behaviour, and ethics.

Ultimately, multi-agentic systems offer not only technical novelty but a radical reconfiguration of the social contract between humans and machines. Designing for this new relational space requires UX frameworks that are sensitive to autonomy, ambiguity, and affect; and that promote human dignity, clarity, and empowerment.






\subsection{Design Infrastructure and Research Methods}
While agentic frameworks are unlocking new capabilities for developers and end-users, the tools and methods available to HCI researchers will need to evolve to keep pace.

First, designing effective multi-agentic interactions demands a rethinking of design principles and interaction patterns. Established patterns in HCI, such as direct manipulation or turn-based dialogue, may fall short when control is shared or emergent. Researchers must identify new patterns for negotiation, escalation, consensus-building, and multi-party conversation management. These patterns can guide the design of systems that are not only usable but also legible and governable.

There is a need for theoretical contributions that ground design in robust conceptual models. Current work on agency, intent, and control in HCI can be extended to account for systems with overlapping and nested goals, asymmetric knowledge, and inter-agent dynamics. Theory-building efforts can help define what meaningful interaction, trust, and oversight look like in the context of multi-agentic systems.

In parallel, HCI researchers must innovate in evaluation methodology. Traditional usability testing may not capture the probabilistic, emergent or asynchronous nature of multi-agentic behaviour. New instruments are needed to assess concepts such as alignment, collective reliability, emotional response to group dynamics, and user confidence in delegation. This includes developing new survey instruments, behavioural protocols, and real-time monitoring tools.

Finally, there is an opportunity to build design guidelines and best practices for managing the inherent complexity of multi-agentic systems. These guidelines could address modularity, explainability, intervention design, and ethical boundaries—enabling practitioners to build systems that are both powerful and responsible.



\section{Conclusion}


In this paper we outline and discuss emerging design challenges in multi-agentic systems, with the perspective of end-users at its core. We emphasise challenges related to the user’s role within multi-agentic systems, the parallel nature of these systems, their emergent complexities, the need to navigate conflicts, and the formation of appropriate mental models, alongside important considerations for trust and explainability. Moreover, we illustrate these challenges through examples and propose design solutions and risk considerations. Our aim is to spark a broader conversation that brings HCI perspectives to the forefront, as we propose a research agenda that directly addresses the critical need for user-centered design to tackle the challenges of multi-agentic systems.




\bibliographystyle{ACM-Reference-Format}
\bibliography{references_sarah, references_yan, references_yan_zotero,ref_wafa}

\end{document}